\documentstyle[aps,prd,epsf,amssymb]{revtex}

\def \be{\begin{equation}}
\def \ee{\end{equation}}
\def \bea{\begin{eqnarray}}
\def \eea{\end{eqnarray}}

\begin{document}

\draft

\tighten

\title{The close-limit approximation to Neutron Star Collisions}

\author{Gabrielle D. Allen${}^{(1)}$,
        Nils Andersson${}^{(2)}$, 
        Kostas D. Kokkotas${}^{(3)}$,\\
        Pablo Laguna${}^{(4)}$,
        Jorge A. Pullin${}^{(4)}$ and
	Johannes Ruoff${}^{(5)}$}

\address{
${}^{(1)}$ Max Planck Institute for Gravitational Physics\\
The Albert Einstein Institute, D-14473 Potsdam, Germany}

\address{
${}^{(2)}$ Department of Mathematics\\
University of Southampton, Southampton SO17 1BJ, UK }

\address{
${}^{(3)}$ Department of Physics\\
Aristotle University of Thessaloniki,
Thessaloniki 54006, Greece}

\address{
${}^{(4)}$ Center for Gravitational Physics \& Geometry\\
Penn State University, University Park, PA 16802, USA}

\address{
${}^{(5)}$ Institut f\"ur Astronomie und Astrophysik\\ 
Universit\"at T\"ubingen, D-72076 T\"ubingen, Germany}

\maketitle

\begin{abstract}
We develop a close-limit approximation
to the head-on collision of two neutron stars
similar to that used to treat
the merger of black hole binaries . 
This approximation can serve
as a useful benchmark test for future fully nonlinear studies.
For neutron star binaries, the close-limit approximation involves assuming 
that the merged object can be approximated as
a perturbed, stable neutron star during the ring-down 
phase of the coalescence. 
We introduce a prescription for the  construction of initial data
sets, discuss the physical plausibility of the various assumptions involved, 
and briefly investigate the character of the gravitational radiation 
produced during the merger. The numerical results show that several of the
merged objects fluid pulsation modes are excited to a significant level. 
\end{abstract}

\widetext

\section{Introduction}
\label{sec:introduction}

The inspiral and subsequent merger of compact binaries is one of the 
most promising sources of gravitational radiation for the new generation 
of interferometric detectors (LIGO, VIRGO, GEO600, TAMA).
Once these instruments come online, 
we hope to learn much 
about physics in strong gravitational fields and at 
super-nuclear densities. This is an exciting prospect, but 
we need reliable theoretical models
of the relevant gravitational-wave signals to 
dig them out of what is likely to be a significantly noisy data stream.
Hence, a huge effort is presently being made to 
model the late stages of binary inspiral, both for black holes and 
neutron stars.
For the inspiral phase the post-Newtonian approximation scheme has been 
pushed to very high orders \cite{PNpertn}, and the available results are now at
a rather satisfactory level of accuracy. For the final merger of the two
binary companions, our understanding is not yet at a comparable level. 
To completely model the collision of two compact objects, one must
resort to numerical relativity and
fully nonlinear simulations assuming no symmetries. 
This provides a great computational challenge, and truly reliable results
may not be available for several years.

In the last few years, our understanding of black hole collisions
has improved considerably. An important reason for the recent advances
is the parallel development of the numerical approach and approximate
methods. In fact, the so-called close-limit
approximation (in which the late stages of the merger is modeled by 
considering the two black holes as a single perturbed one) is in remarkable
agreement with the numerical simulations \cite{price94,anninos}. 
We thus have a
powerful  benchmark test for the fully numerical schemes, which gives us some 
confidence in the physical picture that emerges. 
Of course, a considerable amount of work remains before the black-hole
problem is studied in complete generality. Most importantly, the calculations
must be generalized to include rotational effects \cite{science}. 

Compared to the collision of two black holes, the merging of two neutron 
stars involves, in addition to geometrodynamics, a non-trivial amount
of ``dirty'' physics associated with the stellar fluid and a strong magnetic 
field . In the initial stages
of collision, a shock wave will be generated; the merged object will heat 
up dramatically; the electromagnetic fields of the two stars will 
become intertwined; and a strong burst of energy may result. 
Even though the details remain to be understood, the 
merger is believed to lead to (most of) the observed cosmological 
gamma-ray  bursts \cite{meszscaros-rees}. 
From a gravitational point of view, this is interesting since it means that
gamma ray bursts should be accompanied by a burst of 
gravitational radiation. Thus, there is a possibility that future gravitational-wave observations will help shed light on the gamma-ray puzzle. 
Anyway, at the later stages of the
merger, several avenues are open. If the merged object is too massive to form a
neutron star, it will collapse to form a black hole. But for an interesting 
range of
masses the collapse may be temporarily halted. A supermassive neutron star can
be supported either by thermal pressure or by rotation \cite{shapiro,cook}. In
these cases, the final collapse will occur only after the object either cools
or spins down. The likely extreme rotation of the merged object is relevant
also for other reasons. A rapidly rotating stellar object can be  both
dynamically and secularly unstable. In the first case, the star  spins so fast
that matter is ejected from the equator. Via the so-called bar-mode
instability, the core will shed some of its mass and a  sufficient amount of
angular momentum for it to become dynamically stable \cite{centrella}.  
Once sufficient angular momentum has been shed to bring the rotation 
rate of the merged object below the dynamical Kepler limit,
secular instabilities
may come into play \cite{cfs}. Based on recent results, we would expect
the instability associated with the merged objects $r$-modes 
to spin the core down to a
period longer than 10 ms in a year  or so \cite{andersson,fm,aks,lom}. 
The merged
object should, in fact, evolve almost exactly like a newly born neutron star
(up to the eventual collapse in the case of supermassive configurations).
Hence, the  formation of a massive neutron star
via merger of two less massive ones could be followed by a 
detectable \cite{owen,brady} gravitational wave signal due to the $r$-modes.  

Clearly, it is a very difficult
task to model all the aspects of the merger  
in general relativity. On the other hand,
the payoff
from detecting and understanding these events is enormous. 
Neutron star mergers provide unique laboratories for physics
at an extreme level. Given an understanding of the single 
pieces of the merger puzzle, we can hope to put together an 
accurate picture of the entire event. This paper presents an investigation
into the late stages of binary neutron star mergers. Inspired by the success
of the close-limit approximation for colliding black holes, we
develop an analogous framework for the case of neutron stars. Although 
the close-limit approach to compact object mergers is unlikely to be as 
successful in the case of neutron stars
(see later discussion), our hope is to gain insight into the
character of the gravitational waves that emerge from these events.
Furthermore, the framework that we develop should serve as a useful benchmark
test for future fully general relativistic hydrodynamical simulations.

In the close-limit approach,
the late stages of binary merger are studied by assuming that the initial
configuration can be viewed as (in some sense small) perturbations of the
final object. The background is either a known
analytic or a simple numerical solution. For black hole binaries, the
background is a Schwarzschild or Kerr  spacetime, depending  on whether the
remnant is a slowly or a rapidly rotating black hole. For the situation 
considered in the present paper, head-on neutron star mergers, the background 
can be either a black hole or a non-rotating
relativistic star. The approximation that we will develop is only relevant in
the latter case. The case when a black hole is formed requires a 
study of gravitational collapse, and is beyond our 
present aims. One may argue that the collision of two neutron stars 
should typically lead to an object that is too massive to support 
itself against gravity, and which must therefore collapse to a
black hole. If that were the case, we would be restricted to the presumably
small subset of mergers that involve less massive neutron stars. However,
as mentioned above, supermassive neutron stars can
(at least  temporarily) be supported by both thermal pressure and rotation.
Since  the merged object is likely to heat up to temperatures beyond
$10^{11}$~K and  ought to spin close to the Kepler limit due to conservation
of angular momentum, these effect will be highly  relevant. If the object is 
too
massive  to form a cold non-spinning neutron star, collapse will inevitably 
follow either on the cooling timescale (of the order of 10 s for
cooling due to neutrino  emission) or the spin-down timescale (which depends 
entirely on the mechanism due to which the merged object spins down) . Since 
both of these
time-scales are orders of magnitude longer than the dynamical timescale of the
merged object (typically at the ms level) the immediate aftermath of
merger ought to be attainable within our close-limit approach. 

In the case of colliding black holes, the recent studies show that the 
emerging gravitational waves are dominated by the quasinormal modes of
the final black hole. Intuitively, we expect a similar result for 
merging neutron stars.
The merged object
will pulsate wildly.  This then leads to a  characteristic 
gravitational-wave signal
containing several of the  nonradial modes of oscillation of the newly formed
star. Provided that these modes can be identified in the gravitational-wave
data, their  particulars can be used to put strong constraints on the mass and
the radius (as well as the equation of state) of the star
\cite{prl,astero,apostolatos}. This is an interesting idea, but it requires that
the modes are excited to detectable levels.  Previous simulations indicate
only that the fluid $f$-modes and $p$-modes, 
as well as the gravitational-wave $w$-modes
are  excited when a non-rotating star is perturbed generically 
\cite{aaks,ruoff}. With the
present approximation, we can estimate the  actual level of excitation of
these modes at the late stages of  neutron star merger. This will provide
information that can be used to assess the detectability of these events.

Time evolutions of black hole and neutron star perturbations have
recently received considerable attention. For non-rotating black holes, the
dynamics of perturbations can be investigated via the Zerilli equation 
\cite{zerilli}.
Studies of such problems have led to a much improved
understanding of the dynamics of non-rotating black holes.  
A similar approach has also been  used for perturbed rotating black holes. 
Rotating black-hole perturbations have been studied in the
time-domain by solving the Teukolsky equation \cite{laguna}. For
perturbations of non-rotating neutron stars, there are many possible
formulations, even within a particular choice of gauge.  The results we present
for the 
evolution of initial data representing neutron star head-on mergers were
obtained using two independent codes based on perturbative variables within
the Regge-Wheeler gauge \cite{aaks,ruoff}. 

Given a reliable numerical code to evolve perturbations of a single neutron
star, the close-limit calculation only requires the construction
of relevant initial data. An obvious fundamental difference between close-limit
initial data sets for black hole collisions and those involving neutron
stars is the presence of a horizon. In fact, the success
of the close-limit approach for black holes may to a large extent 
rest on the presence of
the horizon. The reason is the following: The initial data for two colliding
black holes corresponds to rather large perturbations only in the region 
inside the peak (roughly at $r=3M$) of the curvature potential. In effect,
the perturbative scheme is unlikely to be reliable in this region.
However, the bulk of these perturbations never escape to infinity. They are
scattered by the effective potential and subsequently swallowed by the final
black hole. Thus, the anticipated errors in the scheme have a minute
effect on the gravitational radiation that reaches the far zone.
Unfortunately,  the neutron star case is not masked in this way. 
Since one would typically only  include dissipation due to
gravitational radiation in the simplest models, all ``errors'' in our 
initial data
must escape through this channel in the absence of a central horizon. 
Consequently, the outcome of close-limit neutron star
collisions are likely to be  rather sensitive to the particular structure of
the initial data. It should also be pointed out that, 
the fundamental qualitative difference between the two problems makes 
a direct comparison between 
the black hole case and that of neutron stars rather dubious. 

However, a more careful study of the black hole results suggests that the
close-limit approach for neutron stars
may not be a lost cause after all. In close-limit black-hole 
collisions, most
of the energy radiated, if not all,  comes
from quadrupole ($l=2$) quasi-normal mode (QNM) ringing. The only aspect
that distinguishes close-limit initial data from any other data is the fixed
amplitude of the ringing. If we assume that a similar situation arises for
neutron stars, we should focus our attention on the mode-ringing and
ignore the initial burst of radiation. This is natural since
one would expect the initial burst
to be strongly dependent on the characteristics of the initial data. In the
case of stars, this strategy is associated with an unfortunate sacrifice. The
early parts of the gravitational-wave signal from the merger may contain the
rapidly damped $w$-modes \cite{ks,aks}. In the proposed prescription, the inferred
amplitude of these modes may well be unreliable. Actually, one would expect
this  to be the case on physical grounds. The problem of constructing
``astrophysical'' initial data in numerical relativity essentially boils down
to specifying the amount of gravitational waves in the spacetime. In a 
neutron star spacetime, it would be natural
for the initial gravitational waves to escape via the $w$-modes. Hence, an
uncertainty in the  amplitude of these modes is expected, given our poor
understanding of the relevant astrophysical data. Taking the 
suggested
attitude, we do not expect to be able to assess the astrophysical
relevance of the $w$-modes. Instead, we should view the 
neutron star close-limit initial data (and the associated results) 
as a constraint
on the amplitude of the longer lived fluid pulsation modes. 

\section{CLOSE-LIMIT INITIAL DATA}
\label{sec:initial}

\subsection{Stellar Models}

A neutron star model in general relativity is obtained by solving  the
Tolman-Oppenheimer-Volkoff (TOV) system of equations: 
\bea \frac{dm}{dr} &=&
4\,\pi\, r^2\,\rho_{0} \label{eq:mass} \\
\frac{d\nu}{dr} & =& \frac{e^{2\lambda}}{r^2}
\left(m + 4\,\pi\, r^3\,  p_{0}\right)
\label{eq:nu}\\
\frac{dp_{0}}{dr}& =& - (\rho_{0}+p_{0})\,\frac{d\nu}{dr} \,,
\label{eq:press}
\eea
where the spacetime metric is given by 
\be
ds^2 = -e^{2\nu}\,dt^2 + e^{2\lambda}\,d r^2 + r^2\,d\theta^2
+ r^2 \sin^2 \theta\,d \varphi^2\,.
\label{eq:4metric}
\ee
Above, $\rho_{0}$ and $p_{0}$ are the density and pressure, respectively,
and $m$ represents the mass inside radius $r$. It is related to the metric
function $\lambda$ by
\be
e^{-2\lambda} \equiv 1 - \frac{2\,m}{r}\,.
\ee
All quantities are functions
of the radial coordinate $r$ only. The construction of initial data for the 
close-limit approximation involves two steps: First, one needs 
to solve the TOV equations and obtain stellar models for both the colliding 
stars and the final configuration. The second step consists of a suitable
{\it superposition} of colliding stars followed by a {\it subtraction}
of the background star. This yields the perturbations which
are then the focus of the evolution.
With this in mind, we have 
denoted background quantities of the star formed by the merger by the subscript $0$
and perturbations of these quantities by the subscript $1$. The TOV solutions
of the colliding stars will be denoted by an asterisk ($*$).
For simplicity, we will only
consider polytropic equations of state $p_0 = K\,\rho_0\,^{\Gamma}$, where $K$
and $\Gamma$ are the adiabatic constant and index, respectively. The adiabatic
index $\Gamma$ is related to the polytropic index $n$
by $\Gamma = 1+1/n$. In the specific example provided later we use $\Gamma =
2$ ($n = 1$).

\subsection{Initial Perturbations and Constraints}

The construction of astrophysically relevant initial data in 
numerical relativity is an outstanding problem not only
from the  mathematical point of view. In the present
context, we would like the data to represent the initial configuration
for the ring-down phase after a collision.
Without the non-linear evolution that precedes
the ring-down stage, the specification of initial data becomes
a non-trivial ``guess.''
In the 3+1 formulation\cite{adm} of Einstein's equations,
initial data consist of
the spatial metric $g_{ij}$, the extrinsic curvature $K_{ij}$
and the matter fields. These initial
data contain {\it too much freedom}.
Only four out of the twelve components in $(g_{ij},\,K_{ij})$
are fixed by the constraints. The remaining pieces, including
the matter fields, are
freely specifiable and single out the particular
situation under consideration. Furthermore,
it is usually not obvious which four components in
$(g_{ij},\,K_{ij})$ should be obtained from solving
the constraints.

The standard procedure to separate the freely specifiable data from that
fixed by constraints in numerical relativity
is York's conformal approach \cite{york79}. In addition, this method 
provides 
a natural framework for ``superposing" solutions. This is a clear advantage
given our present project. 
For instance, given data $(g_{ij},\,K_{ij})$ for individual
black holes, it is possible to ``add'' these solutions and solve 
the constraints to obtain self-consistent, fully non-linear,
initial data \cite{CL98}.  A similar procedure can be applied
to neutron star binaries.
As mentioned in the previous section, our problem involves
constructing perturbative initial data.
Once the background matter and spacetime is specified, 
the construction of perturbative 
initial data consists of finding solutions that satisfy the Hamiltonian and
momentum constraints to the order required by the perturbative expansion.

As in the non-linear case, close-limit initial data 
is completely characterized, for both black holes and neutron stars,
by the freely specifiable data (data not fixed by the constraints)
and boundary conditions.
That is, both the free data and boundary conditions select from the class of
solutions of the constraints those 
dealing with the superposition of colliding objects.
In this paper, we focus the discussion on simplest possible case: the 
close-limit head-on
collisions of neutron stars that are initially at rest. This means that the
initial extrinsic curvature and matter current density vanish to all orders.
Therefore, the momentum constraints are identically satisfied, and one is only
required to solve the Hamiltonian constraint. More general classes
of neutron star initial data are discussed in \cite{philip}. 
For vanishing extrinsic
curvature, the Hamiltonian constraint reads: 
\be
R = 16\,\pi\,\rho\,,
\label{eq:ham}
\ee
where $R$ is the Ricci scalar constructed from $g_{ij}$ and
$\rho$ is the matter density.

Following York's method, the spatial metric $g_{ij}$ and
matter density $\rho$ are conformally transformed according to:
\bea
g_{ij} &=& \phi^4\,\hat g_{ij}
\label{eq:conf} \\
\rho &=& \phi^{-8}\,\hat\rho\, ,
\label{eq:rhoc}
\eea
where hats denote conformal quantities.
With the above transformation, Eq.~(\ref{eq:ham}) takes the form
\be
8\,\hat\Delta\phi - \hat R\,\phi = -16\,\pi\,\hat\rho\,\phi^{-3}\, ,
\label{eq:hamc}
\ee
where $\hat\Delta \equiv 
\widehat\nabla_i\widehat\nabla^i$ and $\widehat\nabla_i$
denotes covariant differentiation associated with the conformal metric 
$\hat g_{ij}$.

We will assume at this point that metric perturbations only enter
via the conformal factor. That is,
\bea\
\phi &=& \phi_{0}+\phi_{1}
\label{eq:perturb1}\\
\hat g_{ij} &=& \hat g_{ij}^{0}\, .
\label{eq:perturb2}
\eea
However, it is important to realize that {\em this is not a physically
motivated choice}. The reason for choosing data that is spatially
conformally flat (i.e. $\hat g_{ij}^{1} = 0$)  is by no means a necessary condition for
our procedure to work.  In fact, for neutron stars the choice
$\hat g_{ij}^{1} = 0$
could lead to  a suppression in the excitation of $w$-modes \cite{aaks}. 
Still, the
calculations are simplified considerably by this assumption, 
and it allows a better
comparison with the black hole binary case since the initial data for multiple 
black holes is usually constructed assuming conformal flatness
 \cite{bowen-york}.
At one level,
the availability  of 
non-linear evolutions will remove this uncertainty 
since  the free data can then be determined from
the outcome of these evolutions. However, there will still be a 
corresponding choice to make in the specification of initial data for the
nonlinear phase. 

The conformal transformations (\ref{eq:conf}) and (\ref{eq:rhoc}), together with the
perturbative expansions (\ref{eq:perturb1}) and (\ref{eq:perturb2}) yield:
\bea
g_{ij}^{0} &=& \phi_{0}^4\,\hat g_{ij}^{0}\\
g_{ij}^{1} &=& 4\,\phi_{0}^3\,\phi_{1}\,\hat g_{ij}^{0}\\
\rho_{0} &=& \phi^{-8}_{0}\,\hat\rho_{0}\\
\rho_{1} &=& \phi^{-8}_{0}\,\hat\rho_{1}
-8\,\phi^{-9}_{0}\phi_{1}\,\hat\rho_{0}\, .
\eea
With the above perturbative expansions, the Hamiltonian
constraint takes the form
\be
8\,\hat\Delta \phi_{1} - [\hat R\, \phi_{1}
+48\,\pi\, \hat \rho_{0} \,\phi^{-4}_{0}] \phi_{1}
 = -16\,\pi\,\phi^{-3}_{0}\hat\rho_{1} \, .
\label{eq:haml2}
\ee

In summary, given the background ($\phi_{0},\,\hat g_{ij}^{0},\,\hat\rho_{0}$),
close-limit initial perturbations of
head-on collisions of neutron stars initially at rest
consists of only two quantities: the perturbation of the conformal factor,
$\phi_{1}$,
and the perturbation of the conformal matter density, $\hat\rho_{1}$.
In the following, we view the density perturbation $\hat\rho_{1}$ as 
free data, with $\phi_{1}$ obtained by solving the linearized
Hamiltonian constraint Eq.~(\ref{eq:haml2}). Thus, the
 density perturbation $\hat\rho_{1}$
fully characterizes the collision, and we now turn to its specification.

\subsection{A Recipe for Close-limit Superposition of Stellar Models}

The key contribution of the present work is to provide a
recipe for obtaining the density perturbation $\hat\rho_{1}$ from a suitable
superposition of isolated neutron stars.
To represent a ``true'' close-limit approximation, this superposition procedure
must be such  that the perturbations vanish as  the
separation between the stars vanish. In order to
superpose two neutron stars and 
solve Eq.~(\ref{eq:haml2}), it is convenient to perform
a coordinate transformation that brings the 3-metric (\ref{eq:4metric})
into the isotropic (conformally flat) form:
\be
ds^2 = \phi^4_{0} ( d\hat r^2 + \hat r^2\,d\theta^2 
+ \hat r^2 \sin^2 \theta\,d \varphi^2)\,.
\label{eq:cmetric}
\ee
This is accomplished by setting the conformal factor to
\be
\phi_{0} = \left(\frac{r}{\hat r} \right)^{1/2}\,,
\label{eq:cfac}
\ee
and transforming the radial-coordinate according to 
\be
\frac{d\hat r}{dr} = e^{\lambda}\,\frac{\hat r}{r}.
\label{eq:drdr}
\ee
With the metric given by (\ref{eq:cmetric}), Eq.~(\ref{eq:haml2})
can be trivially 
rewritten
as a radial elliptic equation:
\be
\frac{1}{\hat r^2}\frac{d}{d\hat r}\left(\hat r^2\frac{d}{d\hat r}\phi_{1}\right)
- \left[\frac{\l(\l+1)}{\hat r^2} 
+ 6\,\pi\, \hat\rho_{0} \,\phi^{-4}_{0}\right]\,\phi_{1}
= -2\pi\,\phi^{-3}_{0}\,\hat \rho_{1}\, ,
\label{eq:hamr}
\ee
where we have  used a standard decomposition in spherical harmonics:
\bea
\hat\rho(\hat r,\theta,\varphi) & = & \hat\rho_{0}(\hat r) +
                     \sum_{lm} \hat\rho_{1}(\hat r)\,Y_{lm}(\theta,\varphi)
\label{eq:rho1} \label{eq:rho4}\\
\phi(\hat r,\theta,\varphi) & = & \phi_{0}(\hat r) +
                     \sum_{lm} \phi_{1}(\hat r)\,Y_{lm}(\theta,\varphi)
\label{eq:phi1} \,. \label{eq:phi4}
\eea

To proceed, we need to provide a superposition procedure 
from which $\hat\rho_{1}$ can be obtained.
To do this, we make the following
observation: All of the black-hole close-limit initial data sets
considered so far have had not only the appropriate zero-separation limit
but also the correct infinite-separation limit. That is, as the separation of
the black holes increases, the initial data converges to that of two isolated
holes. A similar behavior is in principle desirable for neutron stars.
However, the situation for neutron stars is more complicated because
there are no simple relations between mass, radius and density as
for black holes. Hence, we are forced to use a somewhat ad hoc prescription. 
We neglect tidal deformations etcetera and use 
 the following superposition of density profiles 
of isolated neutron stars:
\be
\hat\rho(\hat r^i) = \hat\rho_*(\hat r^i-\hat\xi^i)+\hat\rho_*(\hat r^i+\hat\xi^i)
-\left[\hat\rho_*(\hat r^i-\hat\xi^i)\,\hat\rho_*(\hat r^i+\hat\xi^i)\right]^{1/2}\,.
\label{eq:rhotot}
\ee
Here $\hat\rho_*$ represents the conformally transformed density profile of the
individual colliding neutron stars displaced
a distance $\hat\xi^i$ in conformal space. For simplicity we are assuming
identical colliding stars. 
It is straightforward to verify that the superposition of 
densities (\ref{eq:rhotot}) 
satisfies both the zero-separation and infinite-separation limits.

Assuming a small displacement (i.e. imposing the close-limit condition),
we find that
\be
\hat\rho_*(\hat r^i\pm\hat\xi^i) = \hat\rho_*(\hat r^i) \pm  \hat\xi^j\,\widehat\nabla_j\,\hat\rho_*(\hat r^i)
         +\frac{1}{2}\,\hat\xi^j\,\hat\xi^k\,\widehat\nabla_j\,
         \widehat\nabla_k\,\hat\rho_*(\hat r^i)\,.
\ee
Thus
\be
\hat\rho = \hat \rho_* 
+ \frac{1}{2}\,(\hat\xi^i\,\widehat\nabla_i\,\hat\rho_*)^2
+ \frac{1}{2}\,\hat\xi^i\,\hat\xi^j\,\widehat\nabla_i\,\widehat\nabla_j\,\hat\rho_* \,,
\ee
so the conformal density perturbation is given by:
\bea
\hat\rho_1 & = &\hat\rho-\hat\rho_0 \nonumber \\
& = &  \hat \rho_* - \hat\rho_0
+ \frac{1}{2}\,(\hat\xi^i\,\widehat\nabla_i\,\hat\rho_*)^2
+ \frac{1}{2}\,\hat\xi^i\,\hat\xi^j\,\widehat\nabla_i\,\widehat\nabla_j\,\hat\rho_*\,.
\label{eq:rho11}
\eea
We choose coordinates such that the centers of the colliding stars lie on
the $z$-axis, so
the off-set vector $\hat\xi^i$ is given by:
\be
\hat\xi^i = \hat\xi\left(\cos{\theta},\,-\frac{1}{\hat r}\,\sin{\theta},\, 0\right)\, .
\label{eq:xi}
\ee
Substitution of (\ref{eq:xi}) into (\ref{eq:rho11}) yields
\be 
\hat\rho_{1}(\hat r,\,\theta,\,\varphi) = \hat\rho_*(\hat r) - \hat\rho_{0}(\hat r)
+ \frac{1}{2}\,\hat\xi^2\,\left[\cos^2{\theta}\,\left(\frac{d}{d\hat
r}\hat\rho_*(\hat r)\right)^2 +\cos^2{\theta}\,\frac{d^2}{d\hat
r^2}\hat\rho_*(\hat r) +\sin^2{\theta}\,\frac{1}{\hat r}\frac{d}{d\hat
r}\hat\rho_*(\hat r)\right] \,. \label{eq:rho_a}
\ee
Making use of $\sqrt{4\,\pi}\,Y_{00} = 1$ and 
$\frac{4}{15}\sqrt{5\,\pi}\,Y_{20} + \frac{1}{3} = \cos^2{\theta}$,
Eq.~(\ref{eq:rho_a}) can be rewritten as:
\bea
\hat\rho_{1}(\hat r,\,\theta,\,\varphi) &=&
\sum_{lm} \hat\rho_{1}(\hat r)\,Y_{lm}(\theta,\,\varphi) \\
&=& \sqrt{4\,\pi}
\bigg[ \hat\rho_*(\hat r) - \hat\rho_{0}(\hat r)
+ \frac{1}{2}\,\hat\xi^2\,\biggl\lbrace\frac{d^2}{d\hat r^2}\hat\rho_*(\hat r)
+ \left(\frac{d}{d\hat r}\hat\rho_*(\hat r)\right)^2
- \frac{4}{\hat r}\frac{d}{d\hat r}\hat\rho_*(\hat r)\biggr\rbrace\bigg] Y_{00}(\theta,\,\varphi) \nonumber \\
&+& \frac{2}{15}\,\sqrt{5\,\pi}\,\hat\xi^2\,
\left[\frac{d^2}{d\hat r^2}\hat\rho_*(\hat r)
+ \left(\frac{d}{d\hat r}\hat\rho_*(\hat r)\right)^2
-\frac{1}{\hat r}\frac{d}{d\hat r}\hat\rho_*(\hat r)\right] Y_{20}(\theta,\,\varphi) \, . 
\label{eq:rho_b}
\eea

From the above results, we deduce that
the conformal density perturbation $\hat\rho_{1}$ has two contributions.
One is a monopole part ($m=0,\, \l= 0$),
\be
\hat\rho_{1} =
\sqrt{4\,\pi}
\bigg[ \hat\rho_* - \hat\rho_{0}
+ \frac{1}{2}\,\hat\xi^2\,\biggl\lbrace\frac{d^2}{d\hat r^2}\hat\rho_*
+ \left(\frac{d}{d\hat r}\hat\rho_*\right)^2
- \frac{4}{\hat r}\frac{d}{d\hat r}\hat\rho_*\biggr\rbrace\bigg]\,.
\label{eq:mono}
\ee
Since a monopole perturbation does not lead to gravitational radiation, we will
not consider this contribution further. The second part is a quadrupole
($m=0,\, \l= 2$) perturbation given by the last term in Eq.~(\ref{eq:rho_b}):
\be
\hat\rho_{1} = 
\frac{2}{15}\,\sqrt{5\,\pi}\,\hat\xi^2\,
\left[\frac{d^2}{d\hat r^2}\hat\rho_*
+ \left(\frac{d}{d\hat r}\hat\rho_*\right)^2
-\frac{1}{\hat r}\frac{d}{d\hat r}\hat\rho_*\right]\,.
\label{eq:rho_final}
\ee
This is the dominant radiative source which should  be used
in Eq.~(\ref{eq:hamr}) to construct initial data. Notice that
the $l=2$ perturbation (\ref{eq:rho_final}) has the correct
limit $\hat\rho_1 \rightarrow 0$ when $\hat\xi \rightarrow 0$
independent of the stellar model used for $\hat\rho_*$.
On the other hand, for the monopole perturbation (\ref{eq:mono})
to vanish in the limit $\hat\xi \rightarrow 0$ we must have 
$\hat\rho_* \rightarrow \hat\rho_0$. We will now discuss 
a specific model in which this is the case. 

\subsection{Defining ``Realistic'' Colliding Star Models}

As we already discussed in the introduction, the final stellar object that is
formed by merger will be rather different from the initial stars.
It will certainly be hotter and most likely spinning more rapidly since
the angular momentum of the inspiral orbit must be conserved. 
In the case we consider here, that of head-on collision of two non-rotating
neutron stars, we obviously need not worry about rotational effects. But we still 
we need to estimate the changes in the equation of state as the stellar 
material 
heats up during the merger.  To do this we must speculate 
what the outcome of the collision may be,
and specify a relation between the final  background
spacetime including $\hat\rho_0$ and the density distribution 
$\hat\rho_*$ of the individual initial stars. 

In order to have
the correct zero- and infinite-separation limits, we must 
consider the relation between the
masses of the colliding stars and the
mass of the background star. A simplifying and to some extent 
reasonable condition is to assume
that the mass lost during the collision is not significant,
so the total mass is approximately conserved. Specifically, 
this implies that the total mass
computed, in the physical space, from the background density $\rho_0$ 
and the total mass obtained from the superposition of densities $\rho_*$ in (\ref{eq:rhotot})
are roughly the same. As we shall see, in the close-limit approximation this
condition implies that properties of the colliding stellar models (for example, the
radius of the stars) depend on the
separation.  

TOV solutions for polytropic equations of state are parameterized
by the central density $\rho_c$, adiabatic constant $K$ and adiabatic index
$\Gamma$. We assume that the collision does not modify
the adiabatic index (and set $\Gamma = 2$ for all models). 
This assumption is consistent with available numerical results \cite{rasio}.
Therefore, the models used to specify $\rho_*$ differ only in central
density and adiabatic constant.  We want these parameters to reflect changes in
the equation of state as the temperature increases, in such a way that we
retain the final, much hotter, star at zero separation. Our method for
constructing polytropic equations of state for the initial stars (in terms of 
$\rho_c$ and $K$) in relation to that of the final star is inspired
by the recent arguments of Shapiro 
\cite{shapiro} for head-on collision of neutron stars from rest at infinity.

The starting point is to notice that TOV solutions
exhibit the following scalings:
\bea
M(\rho_c) &=& \tilde{ M}(\tilde{\rho}_c)\,K^{n/2}
\label{eq:mass_scal}\\
R(\rho_c) &=& \tilde{ R}(\tilde{\rho}_c)\,K^{n/2}
\label{eq:radius_scal}\\
\rho_c &=& \tilde{\rho}_c\,K^{-n}\,,
\label{eq:rho_scal}
\eea
with $M$ and $R$ the total mass and radius of the star.
Above, tildes denote dimensionless quantities.
To use these relations in the close-limit approach, we 
recall that we want our scheme to be valid in two limits: It should lead
to the expected results both in the limit of zero and infinite separation.
To achieve this we
assume that the mass and radius of the colliding 
stars (index $*$) are related to the background star (index $0$) by:
\bea
M_* &=& M_0\,\left(\frac{2-\eta}{2}\right)
\label{eq:mass_c}\\
R_* &=& R_0\,\left(\frac{2-\eta}{2}\right) \,,
\label{eq:radius_c}
\eea
where $\eta$ is a monotonic function of the separation $\xi$ in physical space
and represents
the ``overlap'' of the two stars. Quite naturally, $\eta$ ranges from 0 to 1
with $\eta(\xi\rightarrow 0)=0$ and $\eta(\xi\rightarrow \infty) = 1$. 
This construction ensures that our approximation satisfies the desired limits. 
However, we have as yet no information about the function $\eta$ for
intermediate separations. Such information could be obtained from fully
numerical studies of merging stars. In the present study we make a
natural, albeit quite arbitrary, choice.  We assume that in the close-limit
regime the overlap function is linear in the separation of the
colliding stars and use
\be \eta(\xi) = \frac{\xi}{R_0}\,. \label{eq:eta} \ee
Substitution of 
(\ref{eq:mass_c}) and (\ref{eq:radius_c}) into 
(\ref{eq:mass_scal}) and (\ref{eq:radius_scal}), respectively, now yields
\be
\frac{\tilde{M}_*}{\tilde{M}_0} = \frac{\tilde{R}_*}{\tilde{R}_0} = 
\left(\frac{2-\eta}{2}\right)^{-1}\left(\frac{K_0}{K_*}\right)^{\frac{n}{2}}\,.
\label{eq:rel1}
\ee
Eq.~(\ref{eq:rel1}) implies
\be
\frac{\tilde{M}_*}{\tilde{R}_*} = \frac{\tilde{M}_0}{\tilde{R}_0}\,.
\label{eq:rel2}
\ee
Since the ratio $\tilde{M}/\tilde{R}$ is a monotonic function of 
$\tilde{\rho}_c$, Eq.~(\ref{eq:rel2}) yields 
\be
\frac{\tilde{M}_*}{\tilde{M}_0} = \frac{\tilde{R}_*}{\tilde{R}_0} = 1\,.
\label{eq:rel3}
\ee
Hence, from (\ref{eq:rel1}), (\ref{eq:rel3}) and (\ref{eq:rho_scal}), one obtains
the following relationships between adiabatic constants
and central densities:
\bea
K_* &=& K_0\,\left(\frac{2-\eta}{2}\right)^{2/n}\, 
\label{eq:adiabat_c}\\
\rho^*_c &=& \rho^0_c\,\left(\frac{2-\eta}{2}\right)^{-2}\,.
\label{eq:density_c}
\eea 
In the limit $\eta \to 1$, these scalings reduce to the case of 
two polytropic stars colliding head-on from rest at infinity \cite{shapiro}. 
Here, we will use
(\ref{eq:adiabat_c}) and (\ref{eq:density_c}) to relate the 
polytropic equations of state for the initial and the final stars, 
also for intermediate separations.

It is important to point out that, if the amount of mass lost
during the collision is not negligible conservation of mass is
not given by $M_0 = 2\,M_*$. As is clear from Eq.~(\ref{eq:mass_c}),
 conservation 
of mass would be given by $M_0 = 2\,M_*$ only in the case of infinite separation.
For the situation we consider,
conservation of mass is demonstrated by comparing
the mass $M_0$ computed from $\rho_0$ to that 
from $\rho$ in (\ref{eq:rhotot}). It is the mass obtained from
the superposition rule (\ref{eq:rhotot}) that takes correctly into
account the double counting in the overlap of the density
profiles of the colliding stars.

\section{NUMERICAL RESULTS}
\label{sec:evolutions}

Our numerical procedure to construct initial data consists of first
solving, using a fourth-order
Runge-Kutta integrator, the TOV equations 
(\ref{eq:mass})--(\ref{eq:press}) for the background
and colliding stars, together with
the coordinate transformation (\ref{eq:drdr}).
These solutions are then conformally transformed and used
to compute $\hat\rho_1$. Finally, Eq.~(\ref{eq:hamr}) is solved 
for the perturbation of the conformal factor
$\phi_1$ as a boundary value problem using a standard tridiagonal solver
\cite{nr}.
The boundary conditions for $\phi_{1}$ in Eq.~(\ref{eq:hamr})
are: regularity at the axis, $\phi_{1}|_{\hat r = 0} = 0$, and
asymptotic flatness at infinity,
$\frac{d}{d\hat r}\phi_{1}|_{\hat r = \infty} = 0$.

The main aim of the present paper was to provide a prescription
for close-limit initial data in the case of neutron stars. But, even 
though we will 
not discuss a large sample of numerical evolutions here, it is clearly
appropriate to illustrate typical results obtained for the 
proposed neutron star 
close-limit approximation. We have used two independent numerical codes
 \cite{aaks,ruoff}
to evolve the relevant perturbation quantities from the initial data 
obtained in the previous sections.  These evolution codes have
been well tested, and we have verified that they lead to identical results
in the present case. 

\begin{figure}[fig1]
\leavevmode
\\
\centerline{\epsfxsize=0.8\textwidth \epsfbox{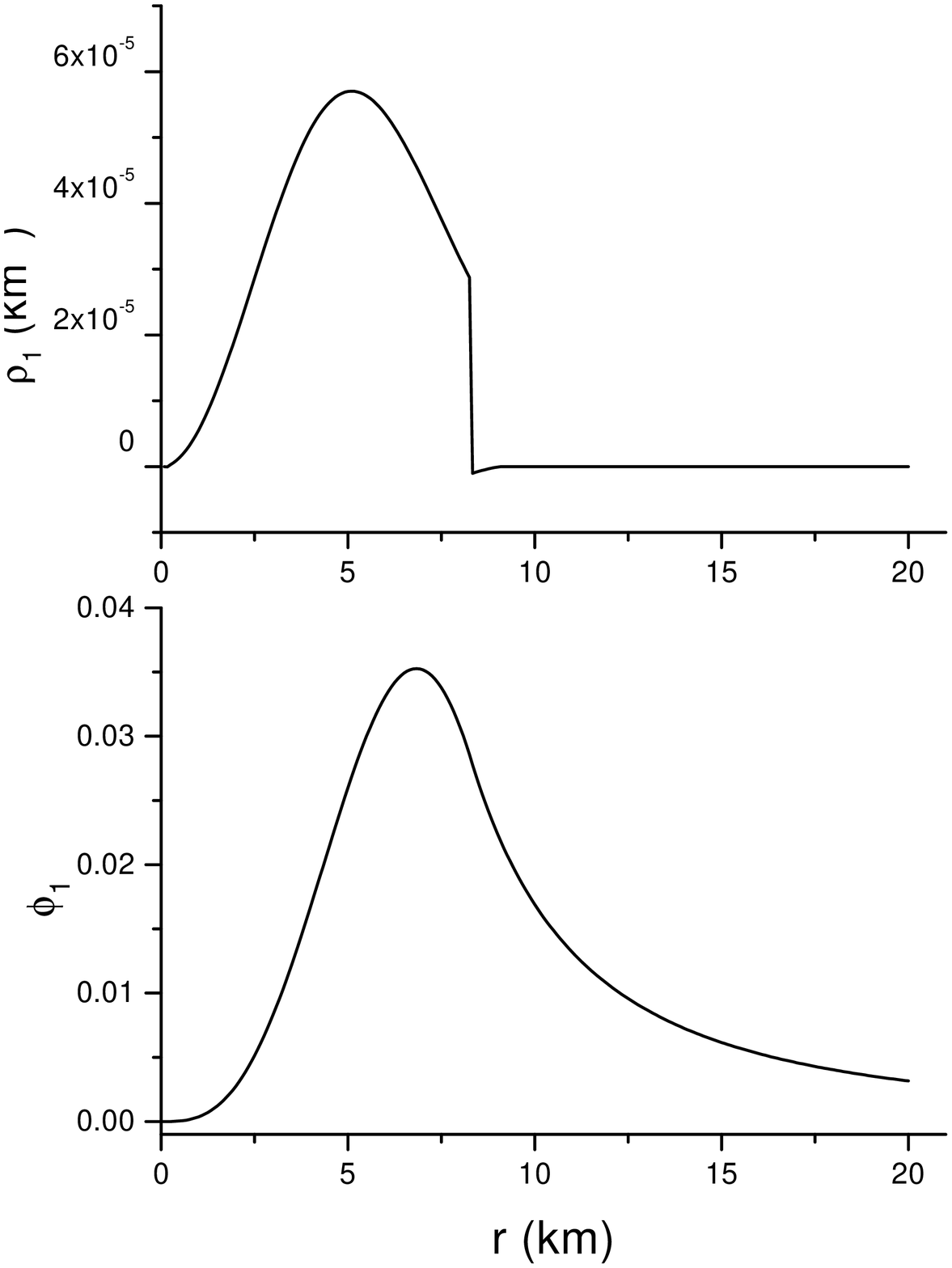}} \
\caption[figure1]{\label{fig1}
Perturbations of the conformal factor $\phi_1$ and
density $\rho_1$ for a 
close-limit collision of equal-mass stars with conformal
separation $\xi = 0.1 R_0$. The TOV parameters for the
background and colliding stars are:
$\rho^0_c = 2.69\times 10^{15}\,\hbox{g/cm}^3$ and  $K_0 = 100\,\hbox{km}^2 $.
For these parameters, the mass and radius of the background star are
$M_0 = 1.24 \,M_\odot$ and $R_0 = 9.0\,\hbox{km}$, respectively. The initial 
colliding stars, which are displaced a distance $0.1R_0$ from the 
center of mass, follow from $\rho^*_c = 2.98\times 10^{15}\,\hbox{g/cm}^3$ 
and  $K_* = 90.25\,\hbox{km}^2 $ and have $M_* = 1.17 \,M_\odot$ and 
$R_0* = 8.58\,\hbox{km}$. }
\end{figure}

\begin{figure}[fig2]
\leavevmode
\\
\centerline{\epsfxsize=0.8\textwidth \epsfbox{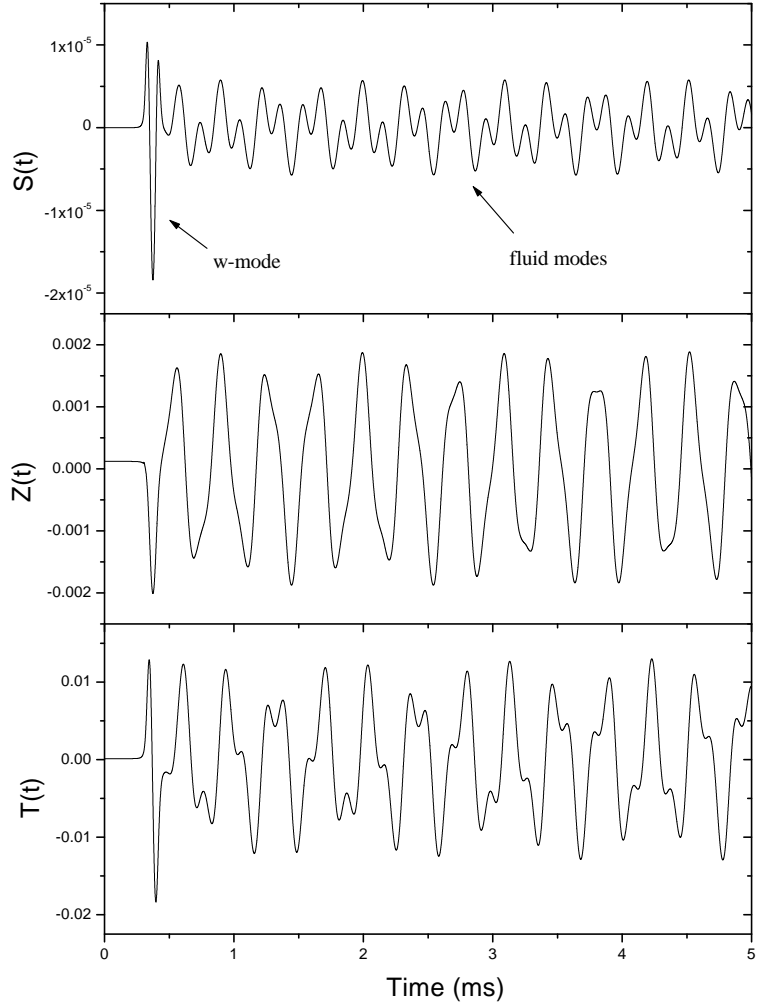}} \
\caption[figure2]{\label{fig2} 
Snapshots of various functions illustrating  for the evolution of the 
initial data shown in Fig.~1. We show the function 
$S$ defined as in \cite{aaks},
the Zerilli function and the perturbed conformal factor 
$T=4r\phi_1/\phi_0$.}
\end{figure}

\begin{figure}[fig3]
\leavevmode
\centerline{\epsfxsize=0.8\textwidth \epsfbox{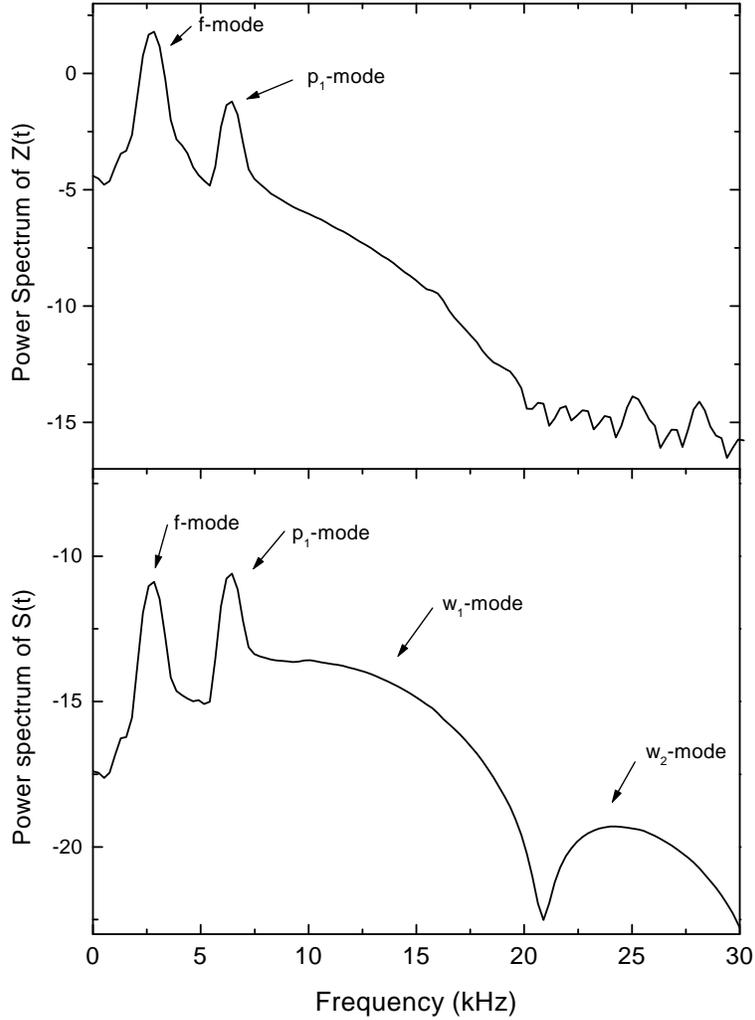}}
\caption[figure3]{\label{fig3} 
The Fourier transform of the waveforms for the function $S$ and
the Zerilli function $Z$ shown in Fig.~2. The two sharp
peaks correspond to the lowest frequency fluid pulsation modes of the 
final configuration.
The first peak belongs to the fundamental $f$-mode while the next one is the 
first of the pressure $p$-modes. It is notable that $w$-modes seem to be 
present in $S$ but not in $Z$. This effect is further discussed in the main 
text.}
\end{figure}

In Figures~\ref{fig1}-\ref{fig3} we show a typical initial data
and  gravitational waves  
resulting from the close-limit 
approximation.  This particular case pertains to
a final stellar configuration with
$\rho^0_c = 2.69\times 10^{15}\,\hbox{g/cm}^3$ and  $K_0 = 100\,\hbox{km}^2 $.
For these parameters, the mass and radius of the background star are
$M_0 = 1.24 \,M_\odot$ and $R_0 = 9.0\,\hbox{km}$, respectively. The initial 
colliding stars, which are displaced a distance $0.1R_0$ from the 
center of mass, follow from $\rho^*_c = 2.98\times 10^{15}\,\hbox{g/cm}^3$ 
and  $K_* = 90.25\,\hbox{km}^2 $ and have $M_* = 1.17 \,M_\odot$ and 
$R_{0*} = 8.58\,\hbox{km}$.
Figure~\ref{fig1} shows the perturbed conformal factor $\phi_1$
and density $\rho_1$. The character of the emerging gravitational
waves is exactly what one would expect: A sharp initial burst followed
by slowly damped oscillations. The long-lived oscillations are associated
with the various fluid pulsation modes of the final configuration.  
That this is the case is clear from Figure~\ref{fig3}, where 
we show the Fourier transform of the waveform in Figure~\ref{fig2}.
In  Figures~\ref{fig2} and \ref{fig3} we can also see that it is 
rather difficult to draw any conclusions about the relevance of the 
short-lived $w$-modes from our sample evolution. While the modes 
are clearly present in the early part of the signal 
if we graph the variable $S$ as defined by Allen et al 
\cite{aaks}, they are absent in the corresponding Zerilli function $Z$ 
which should be a reliable measure of the emerging gravitational
waves. However, at the present time this is perhaps neither here nor there:
As already mentioned, we expect the assumption of conformally flat initial
data to suppress the $w$-modes in a crucial way. Hence, we should 
not attempt to draw any conclusions about their potential 
astrophysical relevance from the present results. 
 
We can, however, meaningfully discuss the longer lived 
fluid modes. 
The various peaks in the spectrum shown in Figure~\ref{fig3}
correspond directly 
to the fluid pulsation modes of the final configuration, the lowest
frequency mode being the fundamental $f$-mode and the next one being
the first of the pressure $p$-modes. This is interesting further evidence
that these modes will be clearly excited whenever a neutron star
is dynamically perturbed, which is highly relevant considering
the recently devised method
for inferring stellar parameters from detected gravitational waves
carrying the mode-signature \cite{astero,apostolatos}. However, 
one must still prove that these modes carry sufficient 
energy to be observable by the new generation of gravitational
wave detectors. To investigate this issue, one should 
study a larger sample of close-limit evolutions and perhaps 
also attempt a comparison to the black-hole case. Work along these
lines is currently in progress, and we hope to report on it
soon. 

\section{Final Comments}
\label{sec:conclusions}

We have developed a framework for studying merging
neutron stars using perturbation theory. Specifically, we have devised
suitable initial data for the very late stages of binary merger, 
when the two merged stars can be considered as a final 
configuration plus perturbations. This ``close-limit'' approximation
is analogous to the one that has provided surprisingly accurate 
results in the case of colliding black holes. However, as we have 
discussed in some detail, it  is not straightforward
 to devise a similar approximation in 
the case of neutron stars. Our chosen 
scheme respects some of the required physical constraints. It has
the correct limits at infinite and zero separation of the 
two stars. Furthermore, we have tried to model the changes in 
the equation of state brought about by the merger in a simple, but
seemingly appropriate \cite{shapiro}, way. Still, it must be 
remembered that this is just a first step and that one could
potentially refine the close-limit idea considerably.  

Even though it is clear that the close-limit approach to neutron star
collisions has severe limitations (it will certainly never completely
replace fully nonlinear general relativistic hydrodynamics simulations) 
we think that it
can prove to be of considerable use. On a technical level, it should be 
rather straightforward to use our initial data sets, combined with the
perturbation evolutions, as benchmark tests for fully nonlinear
evolutions in numerical relativity taking account of the detailed
fluid dynamics. This would in fact be a very useful test for a nonlinear
simulation since it would ascertain that the detailed dynamics
associated with the stars various pulsation modes could be 
resolved. Also, one could clearly use the perturbation equations
to evolve any neutron star process at the late stages, thus saving valuable
computing time. 
Furthermore, it seems likely that we can
learn some important physics from our results. Evolutions from all 
the close-limit data sets that we have so far constructed show
that the fluid pulsation modes of the final star are 
excited to a significant level. As far as the potential 
excitation of the  gravitational $w$-modes
are concerned, our present understanding is far from satisfactory, but
as we have pointed out, an investigation of this issue likely
requires a true understanding of ``astrophysical'' initial data and
a relaxation of the standard assumption of conformal flatness.

In conclusion, 
it is worth pointing out that our close-limit framework
can be extended to  more general situations. 
In the present study we chose to restrict ourselves to 
the head-on collision of two stars that are initially at rest. 
These assumptions can conveniently be relaxed to allow the stars
to have initial momentum. A generalization to the physically relevant
case of slow rotation also seems possible. Work in these directions
is in progress. 

\section*{Acknowledgments}

This work was partially supported by
NATO grants CRG960260 and CRG971092, as well as  
NSF grants PHY9357219, PHY9407194, 
PHY9423950, and PHY9800973.
JP also acknowledges support from the Pennsylvania State University, 
the Eberly Family Research Fund at Penn State, and the John S. Guggenheim
Foundation.

\bibliographystyle{unsrt}

\end{document}